\documentclass[11pt]{article}

\usepackage[T1]{fontenc}
\usepackage[utf8]{inputenc}
\usepackage{amscd,amsmath,amsfonts,amssymb,amsthm}
\usepackage{bm}
\usepackage[mathcal]{euscript}
\usepackage{mathtools}
\usepackage[output-decimal-marker={.}]{siunitx}
\usepackage[only,llbracket,rrbracket]{stmaryrd}

\usepackage{booktabs}
\usepackage{caption}
\usepackage{graphicx}
\usepackage[svgnames,dvipsnames,table]{xcolor}
\graphicspath{{./}{./figures/}}

\usepackage[affil-it,auth-sc]{authblk}
\usepackage[top=1.0in,right=1.0in,left=1.0in,bottom=1.75in]{geometry}

\usepackage[numbers,sort&compress]{natbib}

\DeclareMathOperator{\tr}{tr}

\title{Plastically-driven variation of elastic stiffness in green bodies during powder compaction: \\ Part I Experiments and elastoplastic coupling}

\author[1]{L.P. Argani}
\author[1]{D. Misseroni}
\author[1]{A. Piccolroaz}
\author[1]{Z. Vinco}
\author[2]{D. Capuani}
\author[1]{D. Bigoni\footnote{Corresponding author. Phone:\,+39\,0461\,282507; E-mail:\,bigoni@ing.unitn.it; Fax:\,+39\,0461\,282599.}}

\affil[1]{DICAM, University of Trento, via~Mesiano~77, I-38123 Trento, Italy.}
\affil[2]{Department of Architecture, University of Ferrara, Via Quartieri 8, 44100 Ferrara, Italy}

\begin{document}

\maketitle

\begin{abstract}
\noindent
Cold compaction of ceramic powders is driven by plastic strain, during which the elastic stiffness of the material progressively increases from values typical of granular matter to those representative of a fully dense solid. This increase of stiffness strongly affects the mechanical behaviour of the green body and is crucial in the modelling of forming processes for ceramics. A protocol for ultrasonic experimental investigation (via P and S waves transmission) is proposed to quantify the elastic constants (Young modulus and Poisson’s ratio) as functions of the forming pressure. Experimental results performed in uniaxial strain allow for the introduction of laws that describe the variation of the elastic constants during densification. These laws are motivated in terms of elastoplastic coupling through the simulation of an isostatic pressure compaction process of alumina powder. A micromechanical explanation of the stiffening of elastic properties during densification is deferred to Part II of this study.
\end{abstract}

{\it Keywords: Ultrasound; Elasticity of green body}

\section[Introduction]{Introduction}
\label{S_INTRO}

The investigation of the mechanical properties of tablets and ceramic green bodies is important for the pharmaceutical and the ceramic industry, in the former case as dissolubility and related bioavailability of the drug is related to the density reached during pressing, while in the latter case as the control of ceramic pieces (which have to be handled without failure before firing) is crucial in enhancing the production of both traditional and structural ceramics.
Therefore, the mechanical quality of the green bodies has been so far investigated as, first of all, related to density distribution and strength~\cite{baklouti0, kong, ozkan, walker}, while subsequent experimental campaigns have addressed also fracture toughness and the dependence of elastic properties on forming pressure.
In particular, standard tests~\cite{Cunningham}, pulse-echo~\cite{baklouti, baklouti3} and bending resonance~\cite{njiwa} in a long bar, and non-contact ultrasound with air~\cite{carneim, tittmann}, or water~\cite{rob, schi}, have been employed to measure the elasticity of green bodies.

\begin{figure}[!htb]
\centering
\includegraphics[width=15cm,keepaspectratio]{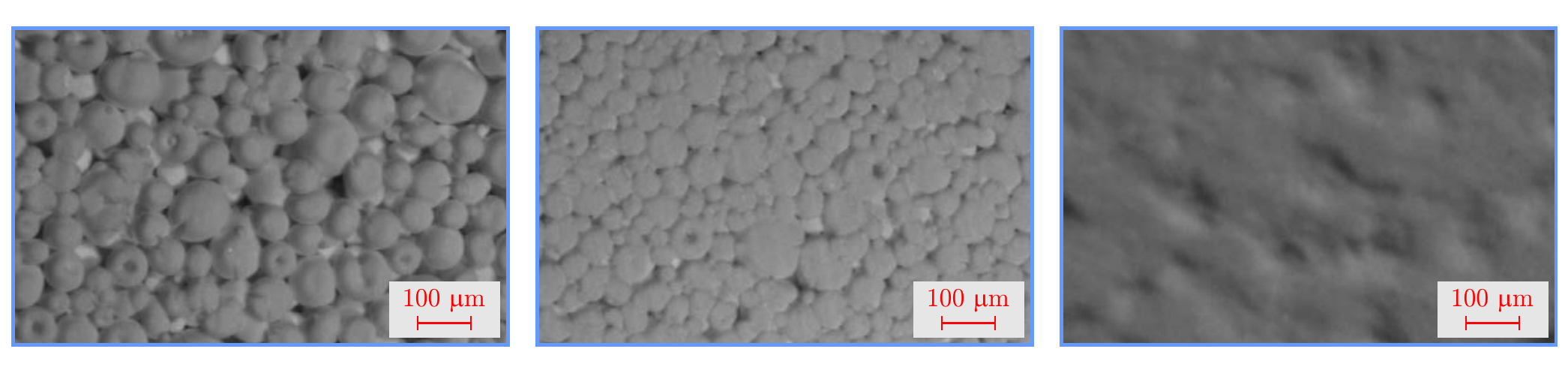}
\caption[Increasing of the contact area between the grains during densification]{
During densification the contact area between the grains increases, as shown by the external surface of cylindrical samples of Martoxid KMS-96 alumina powder compacted under uniaxial strain at different pressures, from left to right: 
1 MPa, 4 MPa, 100 MPa.
Photos have been taken with a Nikon SMZ-800 optical microscope equipped with DSF1i camera head.
}
\label{superfici}
\end{figure}

During cold mechanical densification of a ceramic granulate, the material experiences large plastic strain associated with the enlargement of the contact surfaces between grains (see figure~\ref{superfici}), with a progressive gain in cohesion and elastic stiffening of the material.

In other words, compaction under pressure evolves the elasticity of a powder towards the stiffer elastic response of a fully dense material.
This increase in stiffness is directly related to the development of plastic deformation, so that it can be modelled within the so-called \lq elastoplastic coupling' constitutive framework, which was introduced by T.~Hueckel~\cite{hueck1, hueck2, hueck3} for rock-like materials and used to model and simulate the forming of ceramic powders~\cite{picc1, picc2, stup3, stup1}.
In elastoplastic coupling, a dependence is introduced of elastic parameters on plastic deformation, so that the stiffening of the material during compaction can be modelled.
However, for ceramic granulates the laws governing the dependence of elasticity on plastic strain have been until now only postulated using reasonable assumptions, but never calibrated against direct experiments.
These experiments are one of the objectives addressed in the present article and were carried out by first defining and subsequently applying a testing protocol.
The protocol defines a method of evaluating the elastic parameters of green bodies formed by uniaxial strain compaction through the measurement of ultrasonic P and S wave propagation speeds (Section~\ref{Exp}).

In these experiments the Young modulus and the Poisson's ratio increase after forming by 100\% and 7\% respectively, while a greater variation is observed in the elastic bulk modulus $K$ and in the Lam\'e coefficient $\lambda$.
Three different laws are proposed to describe: (i.) the compaction curves and the variations (ii.) of the wave speeds and (iii.) of the elastic parameters with the forming pressure.
Finally, all the quantities which were found to be nonlinear functions of the forming pressure are shown to increase linearly with density when the latter increases from moderate to high values (but this linearity necessarily does not hold at low densities).
Finally (Section~\ref{modelli}) all the experimental findings are motivated in terms of elastoplastic coupling theory, through the simulation of an isostatic forming process.
The development of a micromechanical model (in which the grains are idealized as elastoplastic circular cylinder or spheres) to explain the variation of the elastic stiffness in green bodies with the forming pressure is deferred to Part~II of this study.

\section[Ultrasound testing of elasticity during powder compaction]{Ultrasound testing of elasticity during powder compaction}
\label{Exp}

\begin{figure}[!htb]
\renewcommand{\figurename}{\footnotesize{Fig.}}
\begin{center}
\includegraphics[width=15cm]{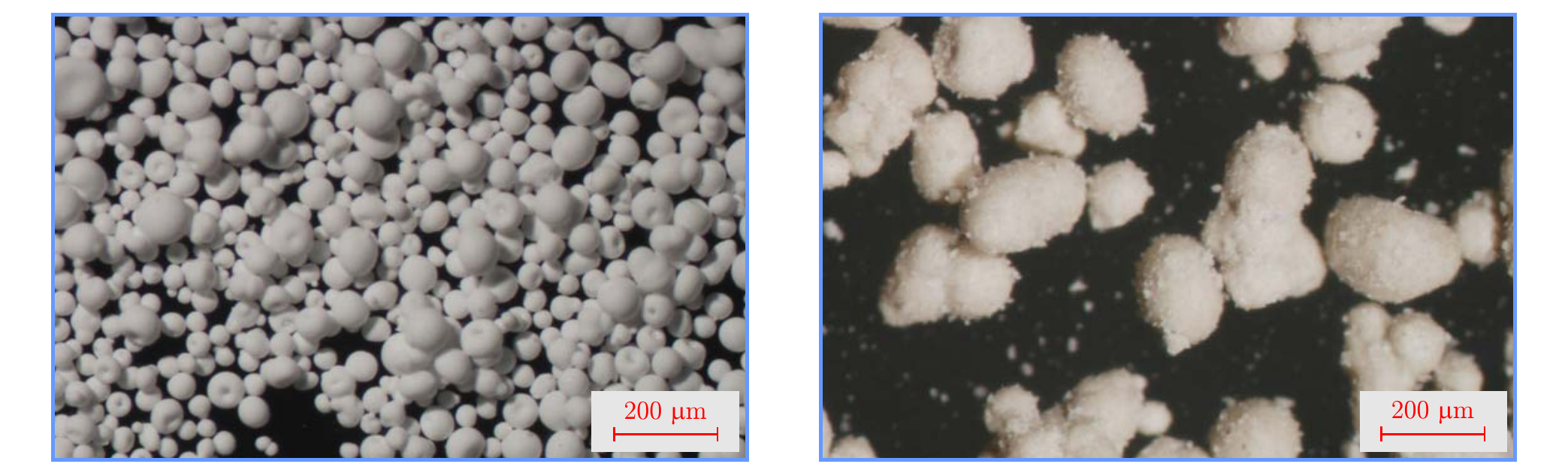}
\caption{\footnotesize Micrograph of the Martoxid KMS-96 powder (left) and of the I14730C powder at 5.5\% water content (right). The photos have been taken with a Nikon SMZ-800 optical microscope equipped with DSF1i camera head. The scale bar is 200 $\mu$m.}
\label{kms2}
\end{center}
\end{figure}

Two different powders have been investigated, one employed for structural and the other for traditional ceramics:

\begin{itemize}

\item a ready-to-press commercial grade, 96\% pure, alumina powder (produced by Albermarle), namely, Martoxid KMS-96.
This powder has particles of \SI{170}{\micro\metre} mean diameter, obtained through spray-drying, figure~\ref{kms2} (left);

\item an aluminium silicate spray dried powder, labeled I14730, tested at two different water contents, namely, $w=5.5\%$ and $w=7.7\%$, corresponding to values used in the industrial forming of traditional ceramics.
The powder has been manufactured by Sacmi S.C. (Imola, Italy) and has the granulometric properties reported in~\cite{bosi}, figure~\ref{kms2} (right).

\end{itemize}

\subsection[Experimental protocol]{Experimental protocol}

What follows is the experimental protocol that was defined for the measurement of the Young modulus and the Poisson's ratio of the ceramic powder for different uniaxial strain values.

\begin{figure}[!htb]
\renewcommand{\figurename}{\footnotesize{Fig.}}
\begin{center}
 \includegraphics[width=8 cm]{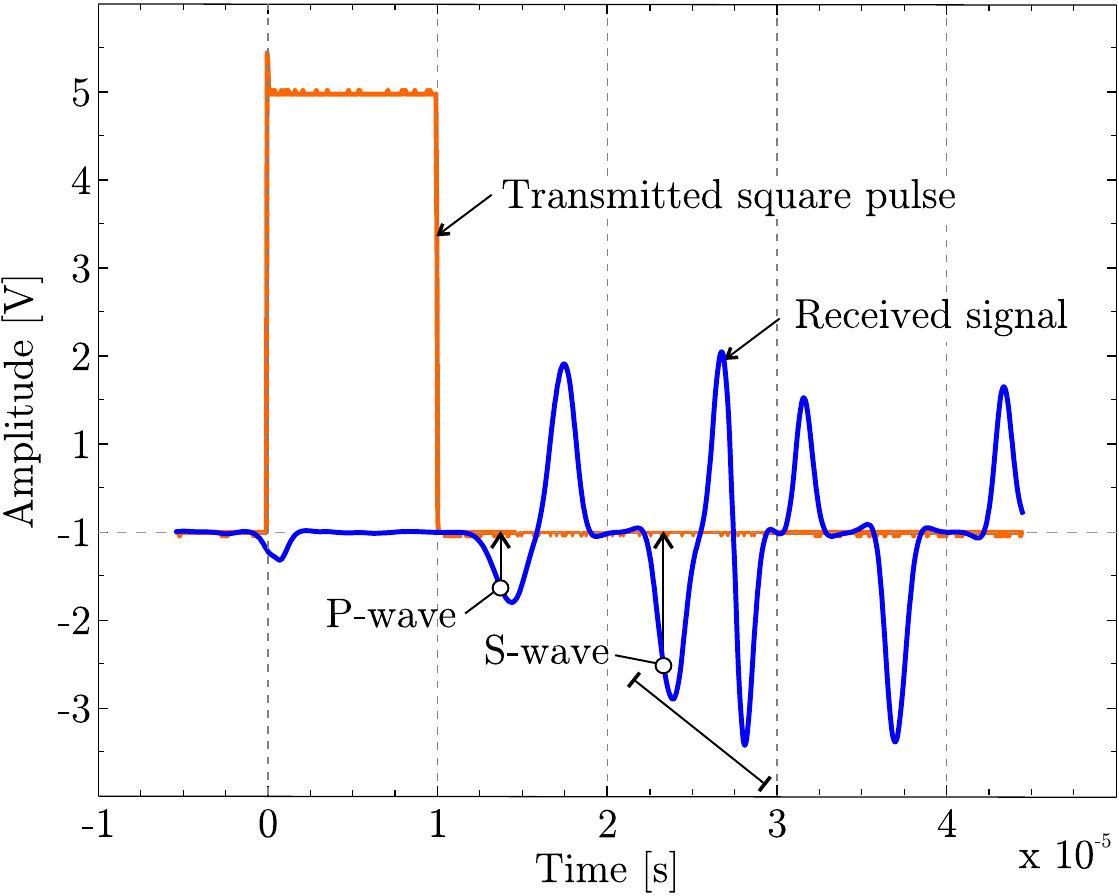}
\caption{\footnotesize An example of filtered signal obtained using two normal incidence shear wave transducers. The signal was measured by a the Digitizer/Oscilloscope and analyzed in Matlab. The transmitted square wave is reported in red and the received wave-form is reported in blue.}
\label{fig:signal}
\end{center}
\end{figure}

\begin{itemize}

\item \SI{10}{\gram} of powder was weighed (using an Orma EB~200 scale, sensitivity $\pm \SI{0.0001}{\gram}$) and was then poured into a stainless steel, cylindrical, single action die (\SI{30}{\milli\metre} diameter) and then settled using vibrations (for~\SI{20}{\second} at~\SI{45}{\hertz}).
The mould was cleaned, before each compaction, with an ethanol soaked cloth.

\item Various disk-shaped specimens were formed by uniaxial strain of the above powder in the mentioned mould to different loading pressures.
This was achieved by imposing a constant rate displacement of~\SI{0.05}{\milli\metre.\second^{-1}} on the top of the mould, using a~\SI{100}{\kilo\newton} electromechanical universal testing machine (Beta~100 from Messphysik Materials Testing); the applied loads were measured with a TC4 load cell from AEP transducers Italy (\SI{100}{\kilo\newton} maximum load); the displacements of the crosshead of the testing machine was measured using a PY-2-F-010-S01M external displacement transducer from Gefran Italy; data were acquired with a NI CompactDAQ system interfaced with Labview~2013, from National Instruments.

\item The thickness of each disk-shaped specimen was estimated using a Palmer caliper (from Mitutoyo, sensitivity $\pm \SI{0.001}{\milli\metre}$), as the mean value of three measures taken at different points, according to the UNI EN~725-10:2008 (Advanced technical ceramics - Methods of test for ceramic powders - Part~10: Determination of compaction properties).
The geometrical density of the green body was then calculated as the ratio between weight and volume of the disk-shaped specimens.
At this stage, the density/pressure response of the material can be plotted (figure~\ref{kms3}).

\item The upper and lower surfaces of the disk-shaped specimens were covered with a layer of either pressure (Sonotech ultragel II ultrasonic couplant) or shear gel (Sonotech shear gel ultrasonic couplant), for use with pressure or shear wave transducers, respectively.

\item The transmission mode technique was adopted in the experimental tests and both shear and longitudinal velocity \textit{independent} measurements in the materials were made.
An ultrasonic square wave pulser/receiver (Olympus~5077PR) unit combined with a NI PCI-5152 Digitizer/Oscilloscope was used with two normal incidence shear wave transducers (Olympus Panametrics NDT V151, frequency~\SI{0.5}{\mega\hertz}) and two pressure wave transducers (Olympus A102S, frequency~1 MHz).

Two transducers were used one as transmitters and the other as receivers.
The signals were measured using a NI PCI-5152 Digitizer/Oscilloscope, and a preamplifier (Olympus~5660B) was used between the receiving transducer and the receiver.
In figure~\ref{equipment} the experimental setup is shown.
The transducers have been selected after a series of preliminary tests performed to find the best transducers in a frequency range between 0.05 and 5 MHz.

\begin{figure}[!htb]
\renewcommand{\figurename}{\footnotesize{Fig.}}
\begin{center}
 \includegraphics[width=13 cm]{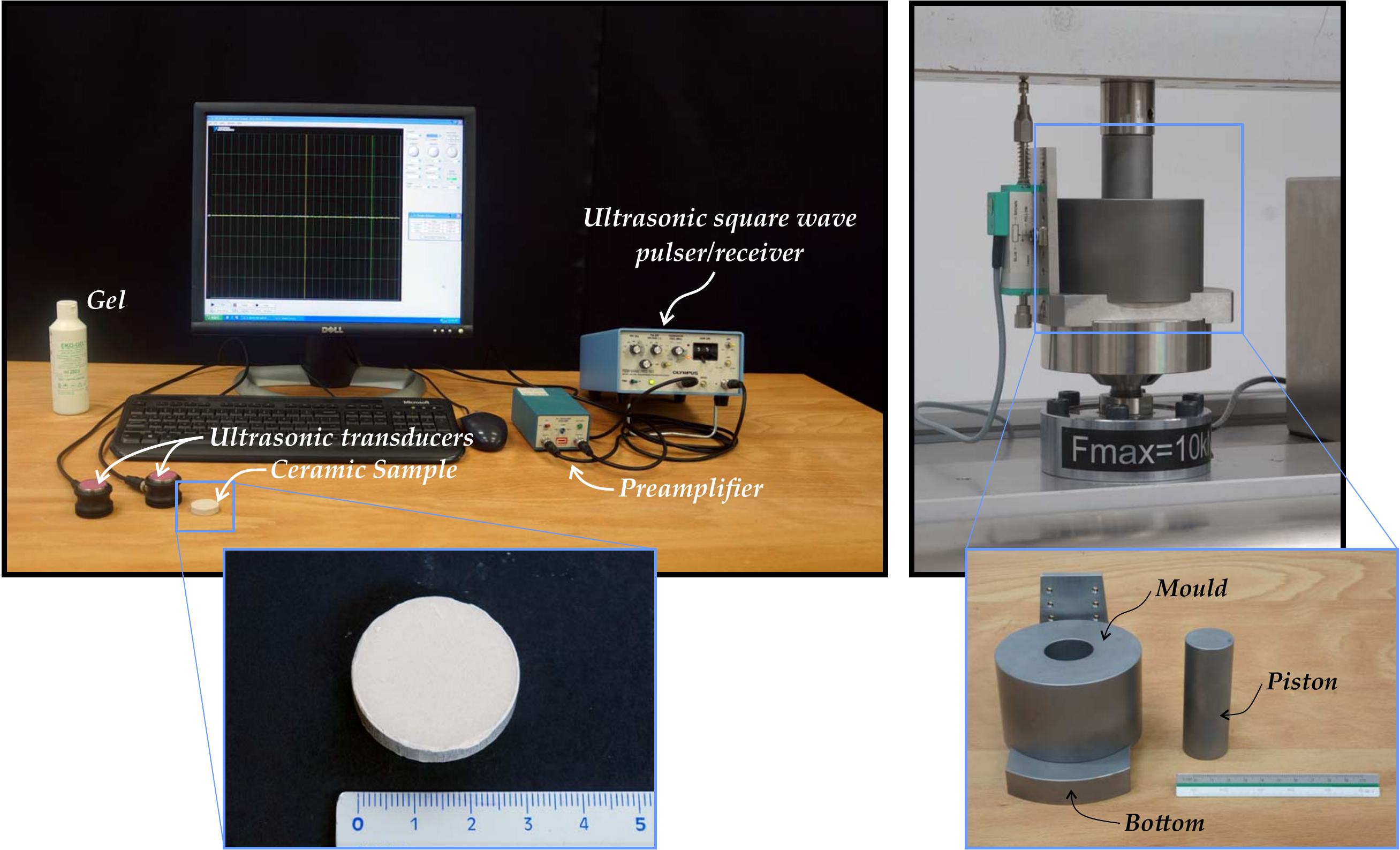}
\caption{\footnotesize The experimental set up for ultrasonic testing. }
\label{equipment}
\end{center}
\end{figure}

The samples (obtained by dry compaction without additives) show low cohesion and are thin, therefore relatively high frequency transducers were chosen, providing a low internal noise even where there are zones of low compaction inside the specimen.
In fact, because frequency and period of a wave are inversely proportional, the lower the transducer frequency the longer is the time interval where noise is present.
For these reasons,~\SI{0.5}{\mega\hertz} transducers for S waves and 1 MHz transducers for P waves were chosen with a receiver gain of~\SI{30}{\decibel}.
On application of these expedients, sufficiently wide and clear signal peaks were obtained without spurious distortions.

\item The analysis of the ultrasonic wave form was divided into two phases. The first phase consisted in observing, directly on the monitor, the NI Scope trace of the received wave for each sample.
If the wave-form was irregular and had a short width it meant that some defects were present and that the sample had to be discarded.
The second phase was implemented on the samples that \lq passed' the first phase and consisted in \lq freezing' and saving the data in a text file for subsequent analysis in Matlab.
From the analysis it was observed that, for times lower than the semi-period associated at the transducer frequency, high level signals are present.
A frequency low-pass filter purifies the signal and shows that the previous mentioned signals are nothing else than spurious noise.

Subsequently, in accordance with UNI EN~583-3 (Non destructive testing - Ultrasonic examination - Transmission technique), the propagation time of the waves was calculated.
The timing point for the arrival of the P wave corresponded with the instant the leading edge of the first peak reached~80\% of its value, which was automatically evaluated 
using a Mathematica routine {\it ad hoc} developed.
For the S wave the timing point was assumed to correspond to a loss in periodicity and increase in signal amplitude, as reported in figure \ref{fig:signal}.

\end{itemize}

\subsection[Experimental results: compaction curves and fitting laws]{Experimental results: compaction curves and fitting laws}

\begin{figure}[!htb]
\centering
\includegraphics[width=17cm,keepaspectratio]{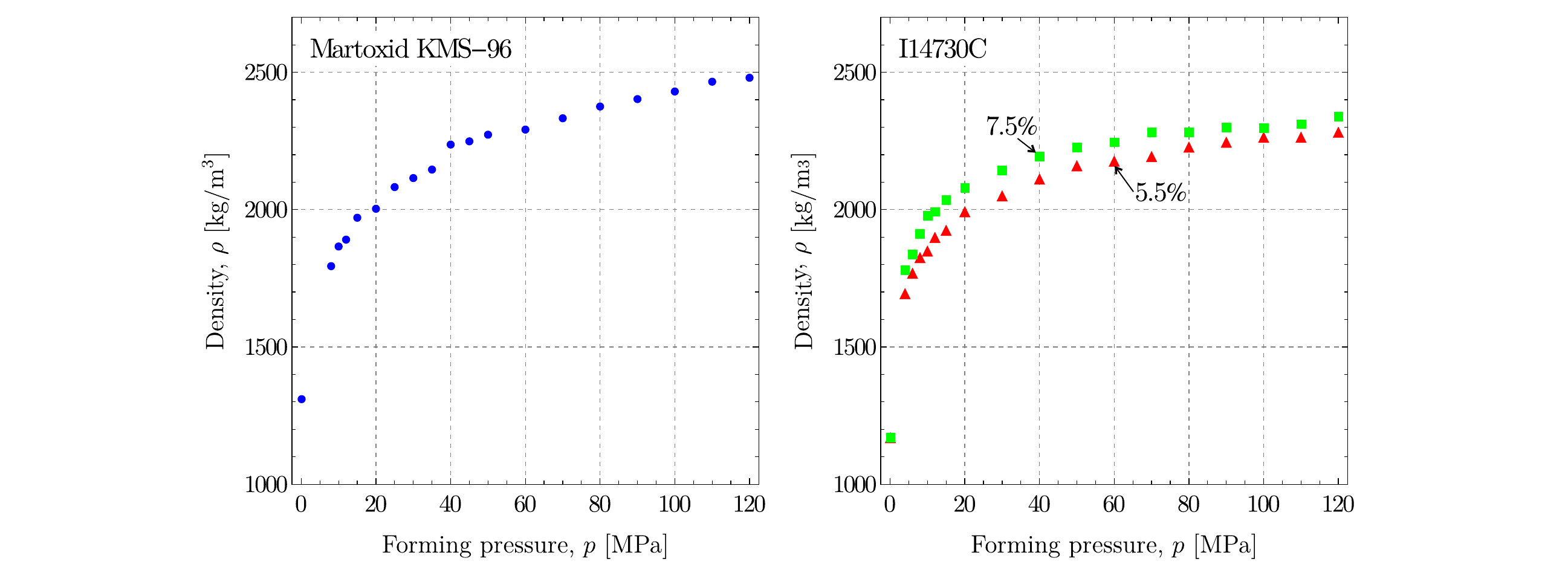}
\caption[Density as a function of the forming pressure]{
Density as a function of the forming pressure for the Martoxid KMS-96 powder (left) and for the I14730C powder at~5.5\% (red/triangles) and~7.5\% (green/squares) water content (right).}
\label{kms3}
\end{figure}

The density is reported in figure~\ref{kms3} as a function of the forming pressure for the two powders (the Martoxid KMS-96 on the left and the I14730C on the right) obtained through uniaxial deformation in a cylindrical, single action stainless steel die (diameter~\SI{30}{\milli\metre}).
Note that the I14730C powder has been tested at~5.5\% (red/triangular spots) and at~7.5\% (green/square spots) water content showing that an increase of the water content yields an increase in the density according to the rule
\begin{equation}
\label{cccc}
\rho=\frac{ \rho_{\textup{dry}} }{1-w} \, ,
\end{equation}
where $\rho$ is the density of the solid and fluid mixture, $\rho_{\textup{dry}}$ is the bulk density of the powder and $w$ the water content.

\paragraph{Fitting laws.}

Several laws have been proposed in the literature to fit compaction curves as those shown in figure~\ref{kms3}.
Three of these fittings are provided and commented below.
All of these three laws are shown to provide a tight interpolation of experimental results.

\begin{itemize}

\item Logarithmic law (LOG): the following law is used within the geotechnical community to model the densification behaviour of geological materials
\begin{equation}
\label{log}
\rho = a \log p + c \, ,
\end{equation}
where $a$ and $c$ are parameters.
The law~\eqref{log} has two advantages, namely, only two parameters are employed and provides the best fitting of results.
However, it has the following disadvantages, namely, it is not defined for $p=0$ and not limited for $p \to \infty$, so that it predicts an unbounded growth of $\rho$ with $p$.

\item Cooper and Eaton exponential law~\cite{cooper} (C\&E):
\begin{equation}
\label{coop}
\rho = \frac{\rho_{0}}{1 - (1 - \rho_{0}/\rho_{\infty})\exp(-b/p)} \, ,
\end{equation}
where $\rho_{0}$ is the initial density, $\rho_{\infty}$ is the upper limit for the density, and $b$ is a parameter related to the slope of the compaction curve.
The law~\eqref{coop} has the advantage that it is limited for $p \to \infty$, but it has the disadvantage that it is not defined for $p=0$ (although the limit as $p \to 0$ is well-defined).

\item \textit{Simple} exponential law (EXP):
\begin{equation}
\label{exp}
\rho = \rho_{\infty} - (\rho_{\infty} - \rho_{0})\exp(-bp) \, ,
\end{equation}
where $\rho_{0}$, $\rho_{\infty}$, and $b$ are parameters with the same meaning of those used in equation~\eqref{coop}.
The law~\eqref{exp} is simple, defined for $p=0$, and limited for $p \to \infty$, but it is not based on a theoretical model (as the other two).

\end{itemize}

\begin{figure}[!htb]
\centering
\includegraphics[width=17cm,keepaspectratio]{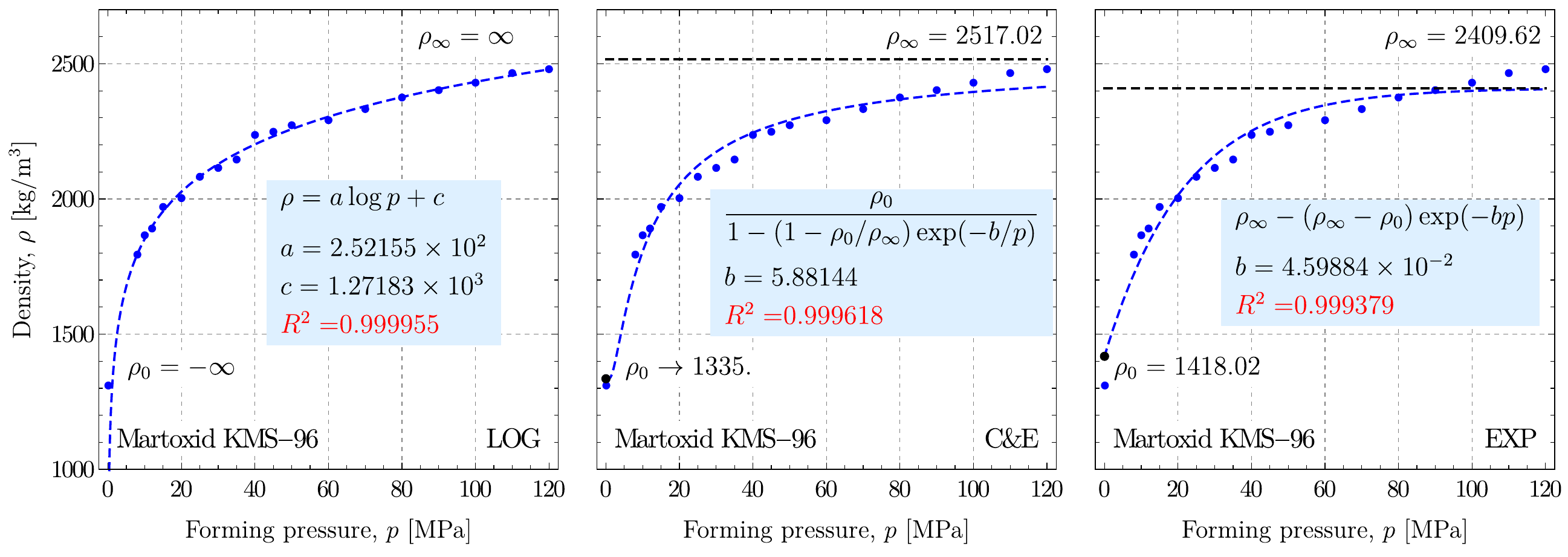}
\includegraphics[width=17cm,keepaspectratio]{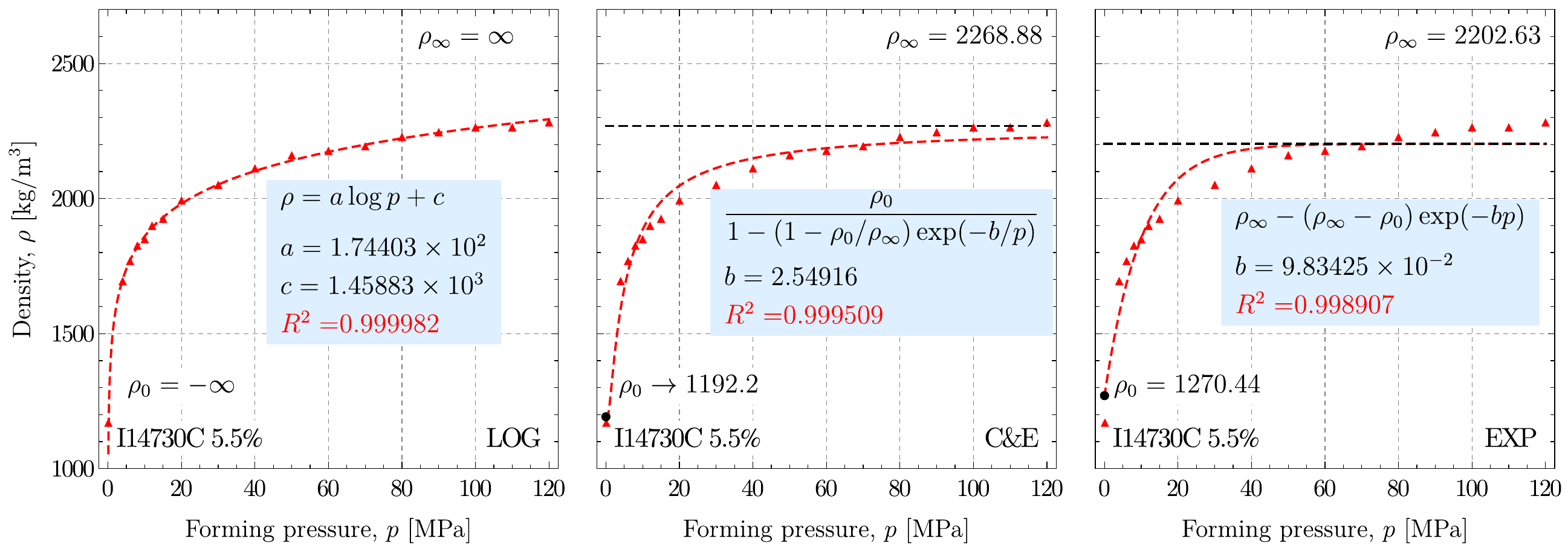}
\includegraphics[width=17cm,keepaspectratio]{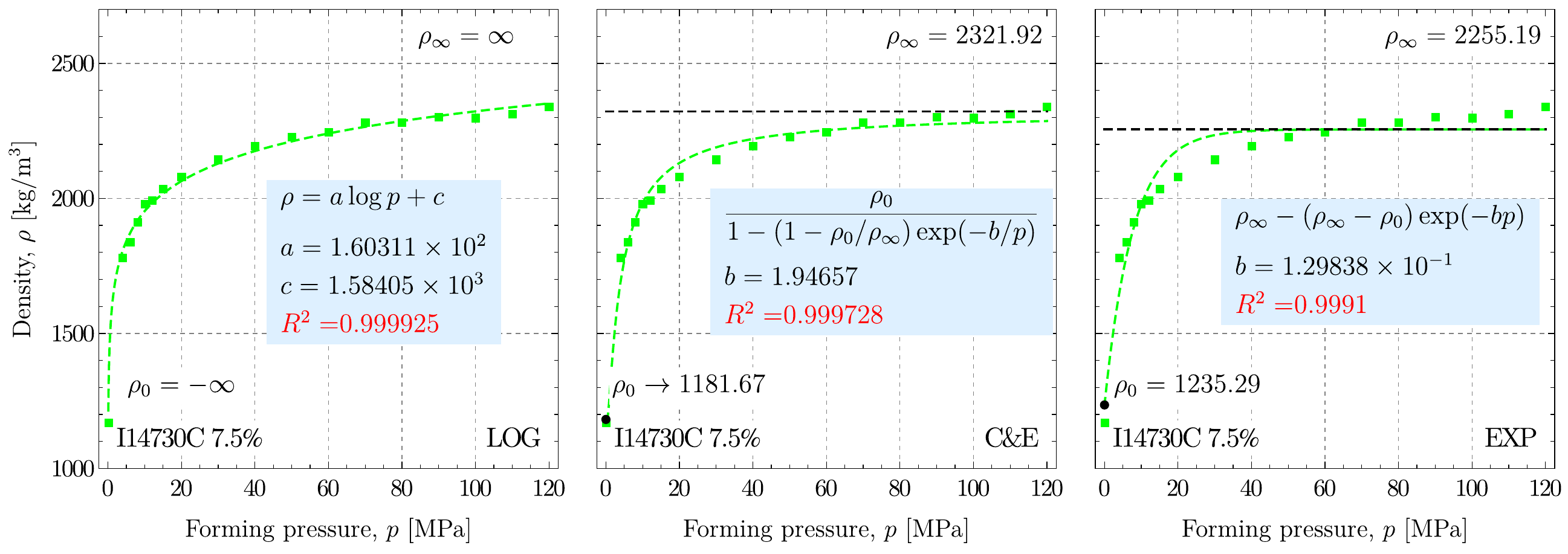}
\caption[Fitting laws for the compaction behaviour of Martoxid KMS-96 and I14730C powders]{
Fitting laws for the compaction behaviour of the Martoxid KMS-96 (upper part), of the I14730C at $w=5.5\%$ (central part) and at $w=7.5\%$ (lower part) powders.
From left to right: Logarithmic law~\eqref{log} (\lq LOG'); Cooper and Eaton law~\eqref{coop} (\lq C\&E'); Exponential law~\eqref{exp} (\lq EXP').
$R^{2}$ denotes the coefficient of determination.}
\label{densityfott}
\end{figure}

The three laws~\eqref{log}--\eqref{exp} have been employed to fit the experimental results reported in figure~\ref{kms3}.
The results are shown in figure~\ref{densityfott}, where it can be deduced that the best fitting is provided by the law~\eqref{log}, but the other two laws are performing reasonably well.

The parameters used to fit the data in figure~\ref{densityfott} were obtained using the FindFit function available within the symbolic computer program Mathematica.
The coefficient of determination $R^{2}$ reported in figure~\ref{densityfott} represents a measure of the goodness-of-fit, so that the better the model fits the data, the closer the value of $R^{2}$ is to 1.
This coefficient is defined as
\begin{equation}
R^{2} = 1 - \frac{ \textup{SS}_{\textup{res}} }{ \textup{SS}_{\textup{T}} } \, ,
\end{equation}
where $\textup{SS}_{\textup{res}}$ is the residual sum of squares and $\textup{SS}_{\textup{T}}$ is the uncorrected total sum of squares.

\subsection[Experimental results: elastic constants]{Experimental results: elastic constants}

\begin{figure}[!htb]
\centering
\includegraphics[width=17cm,keepaspectratio]{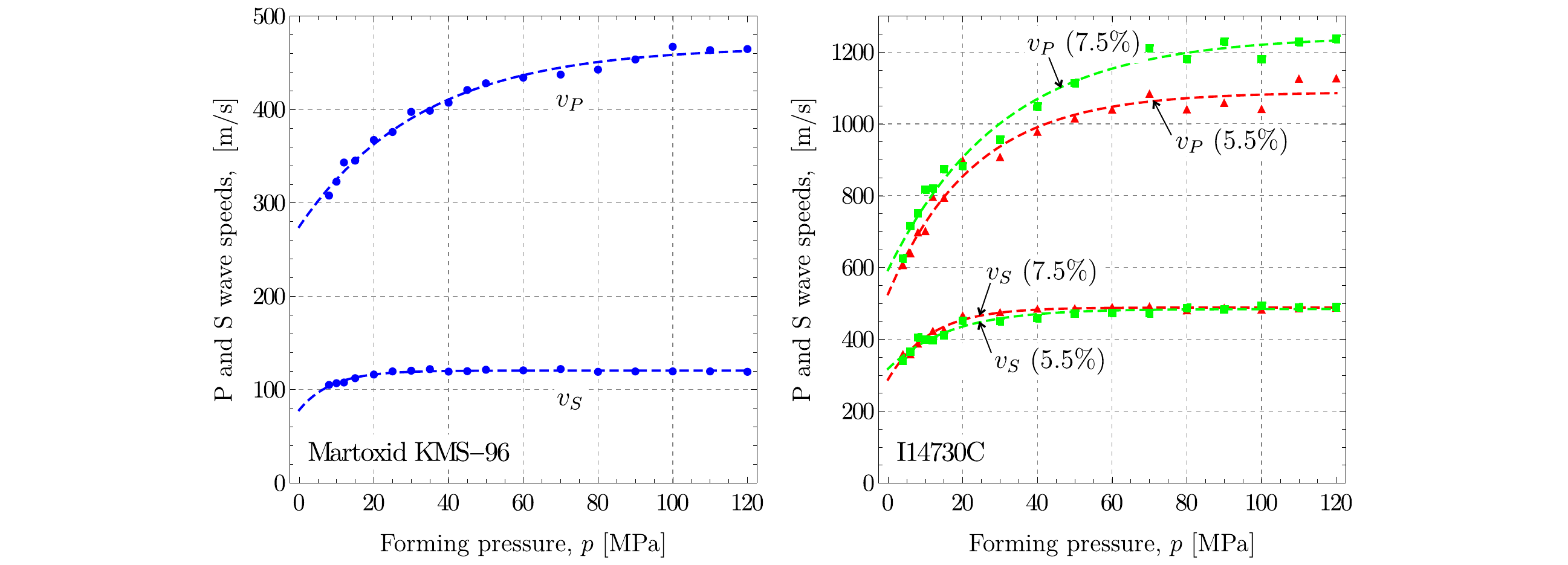}
\caption[Propagation speed of pressure waves and shear waves as a function of the forming pressure]{
Propagation speed of pressure waves and shear waves as a function of the forming pressure for three greens formed from: the Martoxid KMS-96 powder (left) and the I14730C powder at~5.5\% (red/traingular spots) and~7.5\% (green/square spots) water content (right).}
\label{speed}
\end{figure}

The experiments were performed at the \lq Instabilities Lab' of the University of Trento \\ (http://www.ing.unitn.it/dims/ssmg/).
A photo of the experimental setup is reported in figure~\ref{equipment}.
The measured propagation speed of pressure (P) and shear (S) waves is reported in figure~\ref{speed} as a function of the forming pressure for the Martoxid KMS-96 powder (left) and for the I14730C powder at~5.5\% (red/triangular spots) and~7.5\% (green/square spots) water content (right).

The variation of the wave speeds with the forming pressure shown in figure~\ref{speed} has been interpolated with the exponential law~\eqref{exp} and the agreement was found to be pretty good.
Note that the increase in the velocity of P waves with the forming pressure is much more pronounced than the variation of the velocity of S waves.
In both cases the speeds tend to a limit value at increasing $p$, which is consistent with the exponential interpolation law.

\begin{figure}[!htb]
\centering
\includegraphics[width=17cm,keepaspectratio]{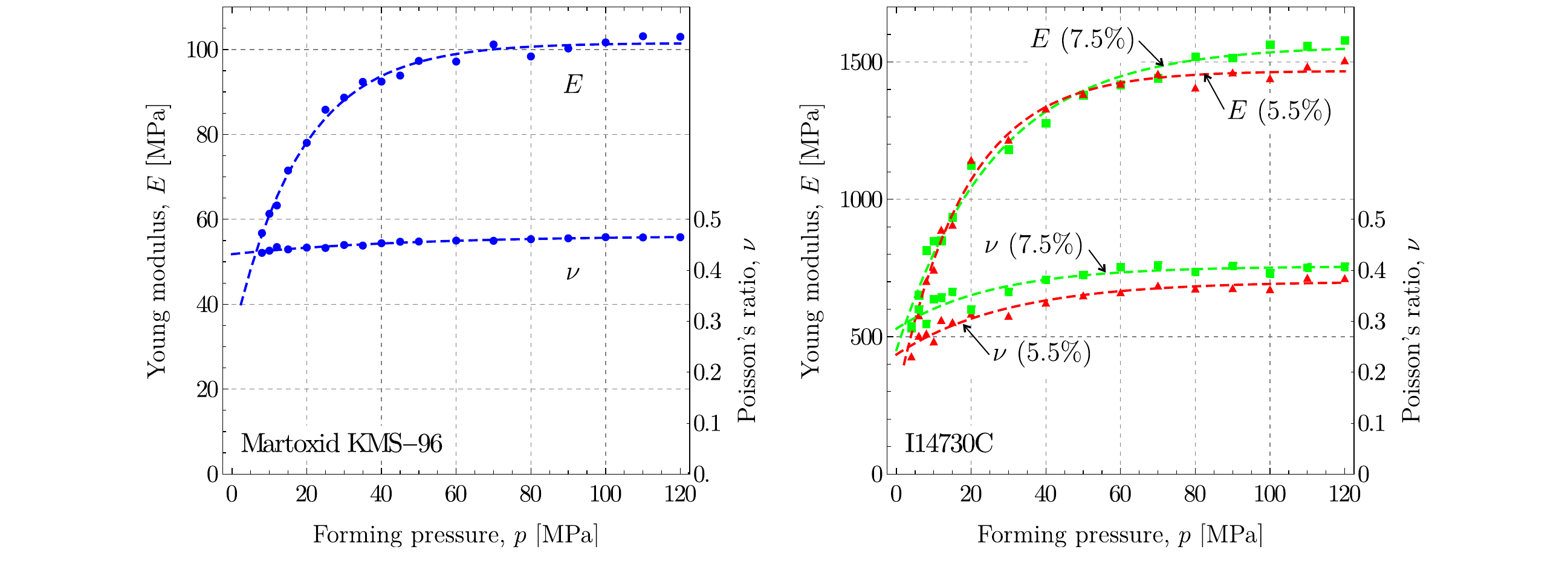}
\caption[Variation of the Young modulus and of the Poisson's ratio with the forming pressure]{
Variation of the Young modulus $E$ and of the Poisson's ratio $\nu$ with the forming pressure for the Martoxid KMS-96 powder (left) and for the I14730C powder at~5.5\% (red/triangular spots) and~7.5\% (green/square spots) water content (right).}
\label{youngnu2}
\end{figure}

\begin{figure}[!htb]
\centering
\includegraphics[width=17cm,keepaspectratio]{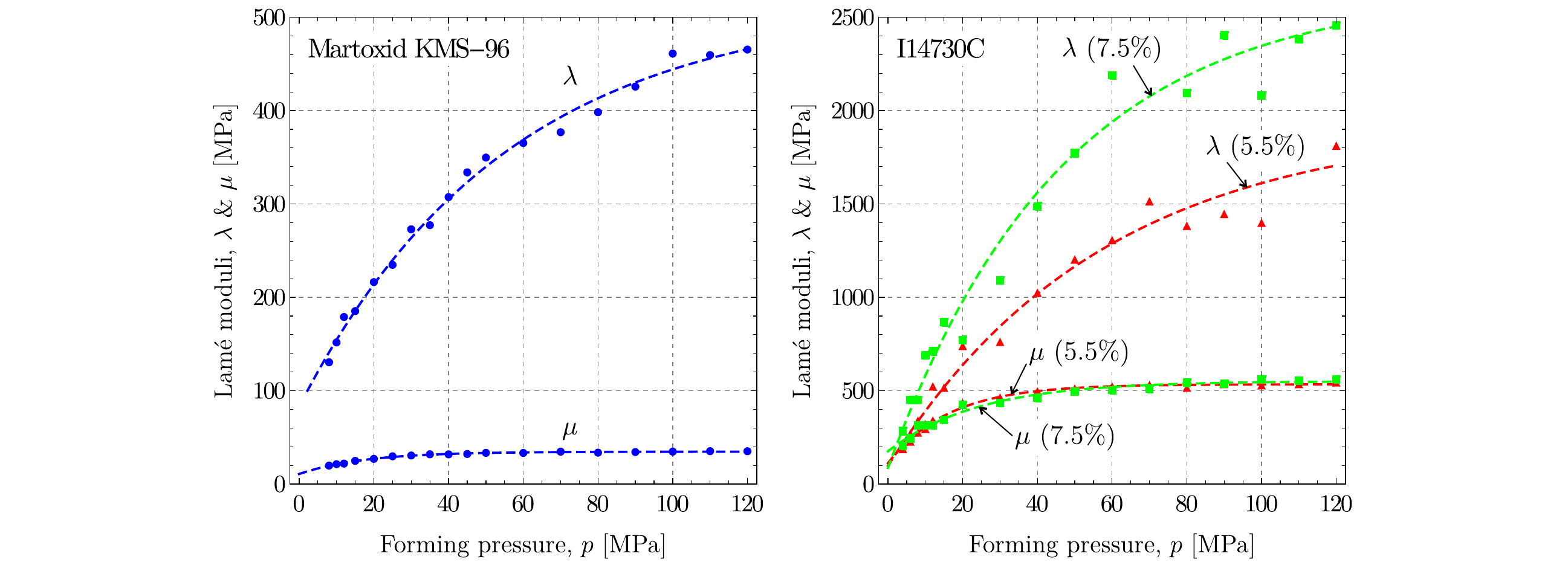}
\caption[Variation of the elastic Lam\'e constants with the forming pressure]{
Variation of the elastic Lam\'e constants $\lambda$ and $\mu$ with the forming pressure for the Martoxid KMS-96 powder (left) and for the I14730C powder at~5.5\% (red/triangular spots) and~7.5\% (green/square spots) water content (right).}
\label{lamu2}
\end{figure}

\begin{figure}[!htb]
\centering
\includegraphics[width=17cm,keepaspectratio]{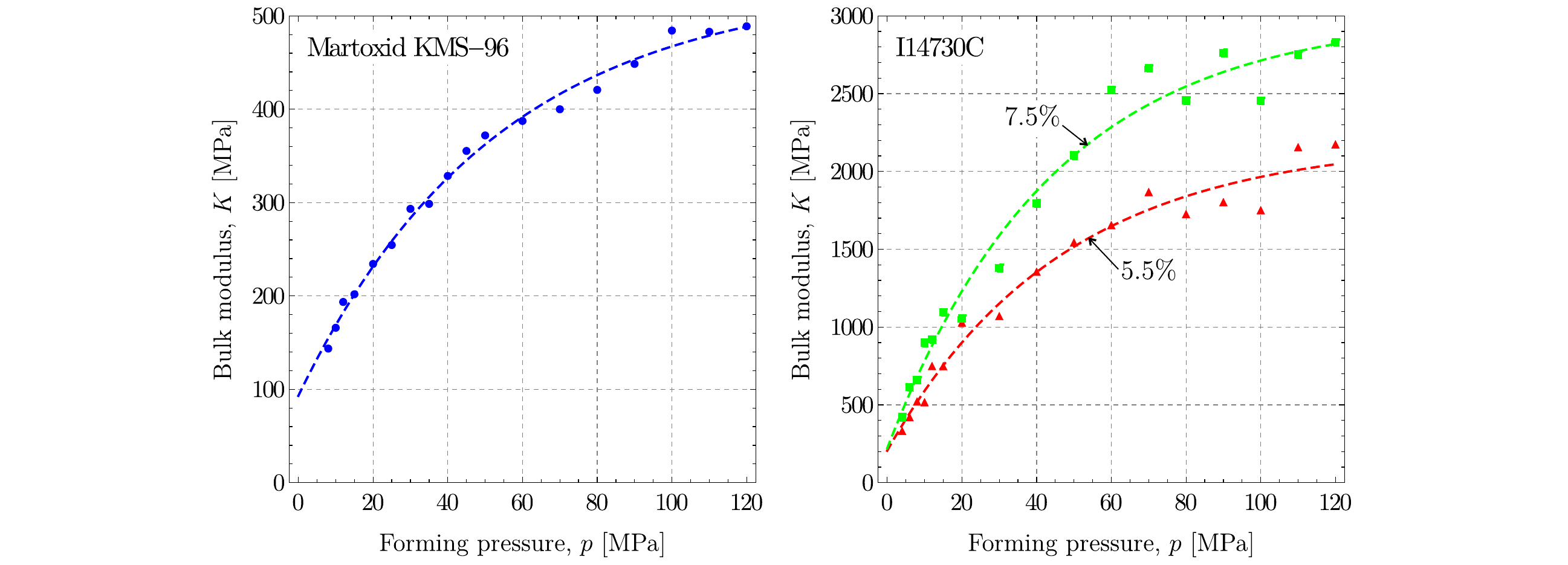}
\caption[Variation of the elastic bulk modulus with the forming pressure]{
Variation of the elastic bulk modulus $K$ with the forming pressure for the Martoxid KMS-96 powder (left) and for the I14730C powder at~5.5\% (red/triangular spots) and~7.5\% (green/square spots) water content (right).}
\label{bulk2}
\end{figure}

The data on wave speeds can be elaborated and transformed in terms of variation of elastic constants as functions of the forming pressure.
In particular, the Young modulus and the Poisson's ratio are reported in figure~\ref{youngnu2}, the Lam\'e moduli $\lambda$ and $\mu$ are reported in figure~\ref{lamu2} and the elastic bulk modulus $K$ in figure~\ref{bulk2}.


All the data reported in Figs.~\ref{youngnu2}--\ref{bulk2} were interpolated with the exponential law~\eqref{exp} and the interpolations, reported in the figures, show a very satisfactory agreement.
All the elastic constants are shown to increase with the forming pressure and eventually reach a limit value.
The increase in the elastic modulus $E$ is approximately of~100\%, in the constant $\lambda$ and in the bulk modulus of the~500\% or more, while the increase in the shear modulus $\mu$ is only of the~80\% and in the Poisson's ratio of a modest~7\% (it should be remembered that the Poisson's ratio lies in the interval (-1,1/2)).
It can therefore be concluded that, while the variation in the Poisson's ratio can be in a first approximation neglected, all the other (dimensional) elastic constants display a non-negligible variation.


\subsection[Variation of the elastic properties with density]{Variation of the elastic properties with density}

\begin{figure}[!htb]
\centering
\includegraphics[width=17cm,keepaspectratio]{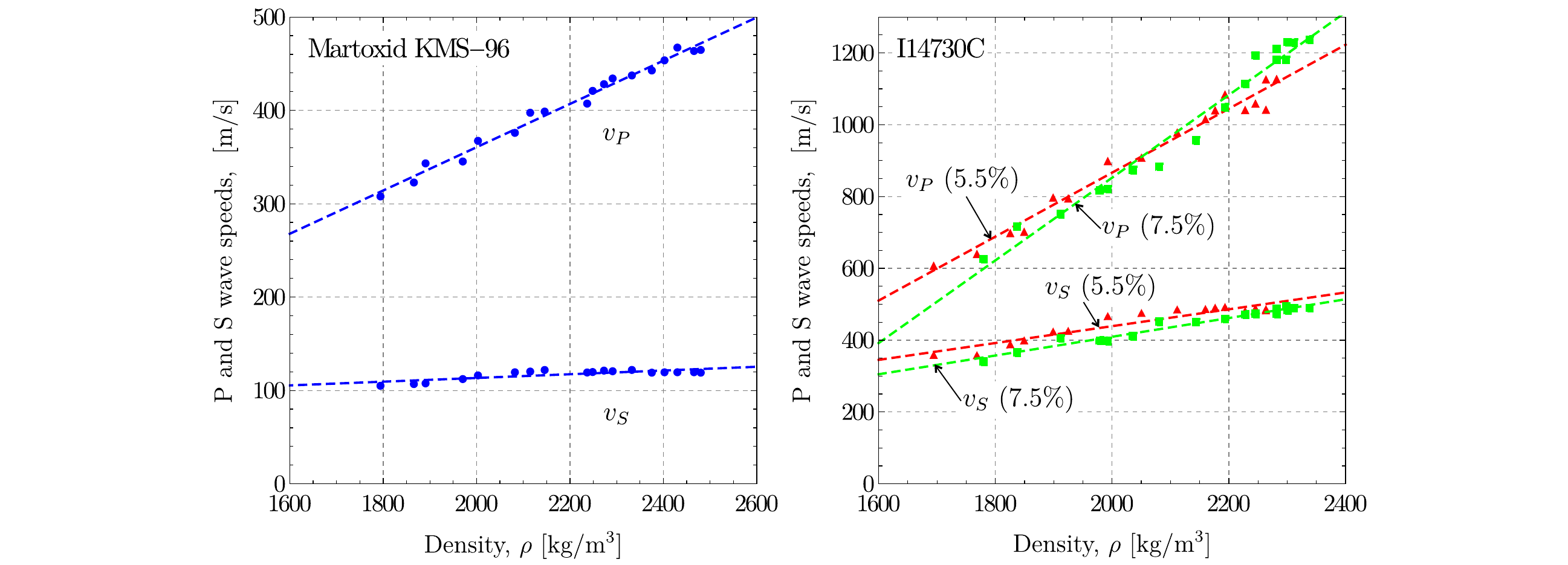}
\caption[Propagation speed of pressure waves and shear waves as a function of the green's density]{
Propagation speed of pressure waves and shear waves as a function of the green's density for the Martoxid KMS-96 powder (left) and the I14730C powder at~5.5\% (red/triangular spots) and~7.5\% (green/square spots) water content (right).}
\label{speedlin2}
\end{figure}

\begin{figure}[!htb]
\centering
\includegraphics[width=17cm,keepaspectratio]{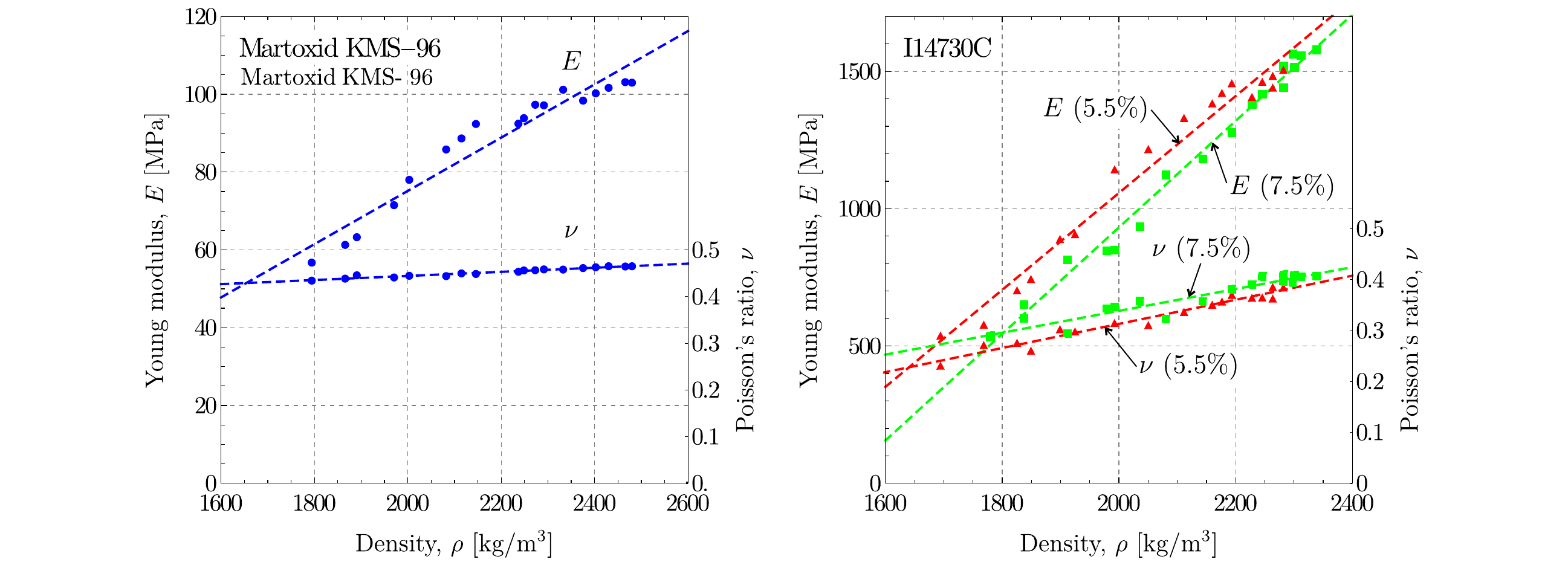}
\caption[Elastic Young modulus and Poisson's ratio as functions of the green's density]{
The elastic Young modulus $E$ and the Poisson's ratio $\nu$ as functions of the green's density for the Martoxid KMS-96 powder (left) and the I14730C powder at~5.5\% (red/triangular spots) and~7.5\% (green/square spots) water content (right).}
\label{youngnulin2}
\end{figure}

\begin{figure}[!htb]
\centering
\includegraphics[width=17cm,keepaspectratio]{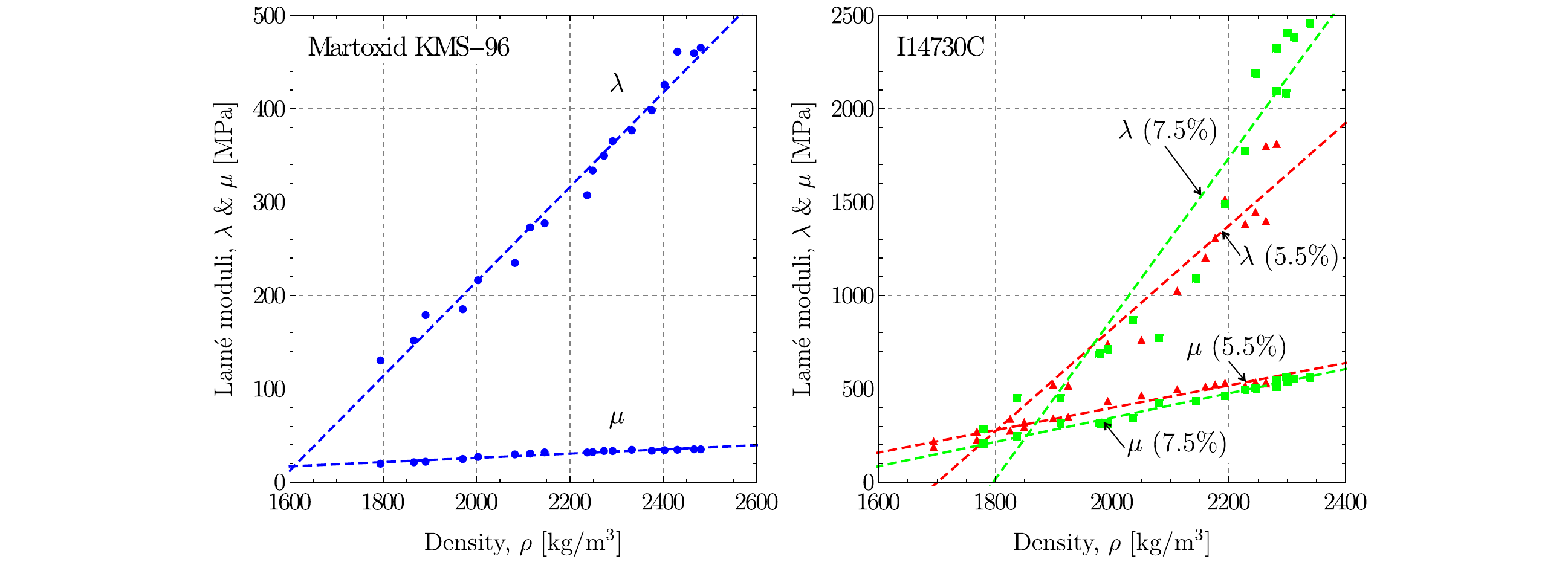}
\caption[Lam\'e elastic constants as functions of the green's density]{
The Lam\'e elastic constants $\lambda$ and $\mu$ as functions of the green's density for the Martoxid KMS-96 powder (left) and the I14730C powder at~5.5\% (red/triangular spots) and~7.5\% (green/square spots) water content (right).}
\label{lamulin2}
\end{figure}

\begin{figure}[!htb]
\centering
\includegraphics[width=17cm,keepaspectratio]{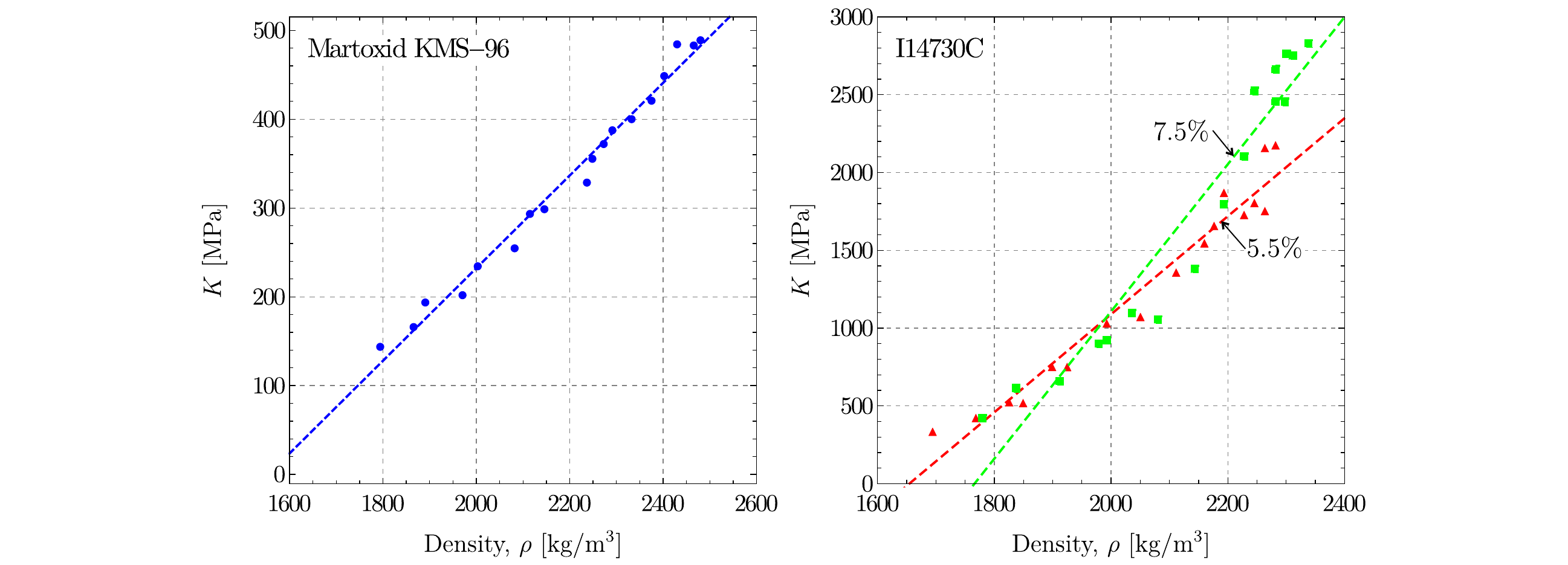}
\caption[Elastic bulk modulus as a function of the green's density]{
The elastic bulk modulus $K$ as a function of the green's density for the Martoxid KMS-96 powder (left) and the I14730C powder at~5.5\% (red/triangular spots) and~7.5\% (green/square spots) water content (right).}
\label{bulklin2}
\end{figure}

In contrast with the theory developed by Kendall et al.~\cite{kendall}, where the speed of P waves is proportional to $\rho^{3/2}$, a linear dependence was found of the P waves velocity
(ranging in their experiments between~\num{500} and~\SI{900}{\metre.\second^{-1}}) on the density of the green~\cite{carneim, kupperman}.
Therefore, the dependence of P and S wave speeds, as well as of all the elastic parameters, on the green density has been investigated and indeed found to be linear in all cases.
In particular, figure~\ref{speedlin2} shows that the dependence on the density of the samples of the P and S wave speeds is linear and the same linearity is visible in Figs.~\ref{youngnulin2}--\ref{bulklin2} reporting about the Young modulus $E$, the Poisson's ratio $\nu$, the shear and bulk moduli $\mu$ and $K$, and the Lam\'e constant $\lambda$.

It is important to emphasize that the linear interpolations shown in Figs.~\ref{youngnulin2}--\ref{bulklin2} hold from medium to high density and cannot be applied at low pressures.
In fact, they would predict \textit{negative} values of the elastic constants at sufficiently low density, which is certainly not possible.
Therefore, for practical use, the linear interpolation laws have to be augmented with a nonlinear law providing a null value of the elastic constants in the limit of null density.
An example of such a law is formulated below for the bulk modulus $K$ reported in figure~\ref{bulklin2}.
An interpolation of the data reported in this figure, relative to the Martoxid KMS-96 powder, valid also for low density is the following
\begin{equation}
\label{wlf}
K(\rho) = s(\rho) f_{1}(\rho) + \bigl( 1 - s(\rho) \bigr) f_{2}(\rho) \, ,
\end{equation}
where
\begin{subequations}
\begin{align}
f_{1}(x) &= \num{0.521499} \, x - \num{810.774}                  \, , \\
f_{2}(x) &= \num{3.46123 e-13} \, x^{\num{4.49153}} + \num{0.01} \, ,
\end{align}
\end{subequations}
and $s(x)$ is the smooth unit-step function defined as
\begin{equation}
s(x) = \frac{1}{2} \left( 1 + \tanh \left( \frac{x-\num{2000}}{100} \right) \right) \, .
\end{equation}

\begin{figure}[!htb]
\centering
\includegraphics[width=17cm,keepaspectratio]{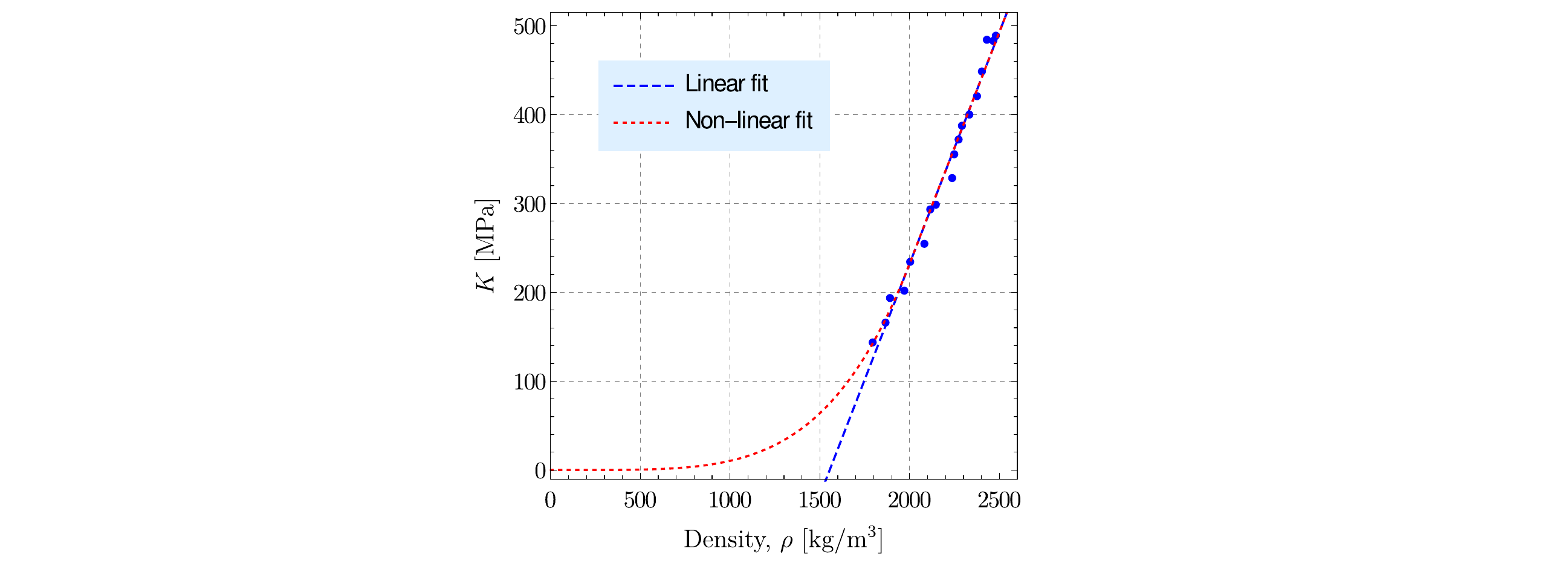}
\caption{Interpolation of the variation of the elastic bulk modulus with density, equation~\eqref{wlf}.
At high density the variation is linear, but has to become nonlinear at low density (otherwise the elastic bulk modulus would assume negative values). }
\label{Krhosmooth}
\end{figure}

The law~\eqref{wlf} is reported in figure~\ref{Krhosmooth} together with the experimental data and shows an excellent fitting.

\section[Variation of the elastic properties with density: effects on elastoplastic coupling]{Variation of the elastic properties with density: effects on elastoplastic coupling}
\label{modelli}

The variation of the elastic properties with density represents a coupling between the elastic and plastic characteristics of the material, so that the irreversible deformation producing the compaction of the green body influences its elastic stiffness.
The elastoplastic coupling is not usually accounted for in the modelling of the plastic flow of metals (see for example~\cite{hill}), but was introduced by T.~Hueckel to describe
the degradation of elastic stiffness observable during inelastic deformation of granular and rock-like materials~\cite{hueck1, hueck2, hueck3}.
In the following the consequences of the dependence of elastic constants on the density of the green are analyzed with a specific example of isostatic pressing of a ceramic granulate.

\subsection[Elastoplastic coupling]{Elastoplastic coupling}

According to the elastoplastic concept~\cite{biiii}, the stress rate is the sum of a \lq standard' contribution (related to the difference between the strain rate $\dot{\bm{\varepsilon}}$
and the plastic part of the strain rate $\dot{\bm{\varepsilon}}_{\textup{p}}$) and a contribution related to the variation of the elastic properties
(represented here by the fourth-order tensor ${\mathcal E}$)
\begin{equation}
\label{couplazzo2}
\dot{\bm{\sigma}} =
\underbrace{
{\mathcal E}[\dot{\bm{\varepsilon}} - \dot{\bm{\varepsilon}}_{\textup{p}}]
}_{\text{\lq standard'~term}}
+
\underbrace{
\frac{\partial {\mathcal E}}{\partial \bm{\varepsilon}_{\textup{p}}}[\bm{\varepsilon} - \bm{\varepsilon}_{\textup{p}}] \dot{\bm{\varepsilon}}_{\textup{p}}
}_{\text{e-p coupling}} \, ,
\end{equation}
so that for an elastic isotropic material
\begin{equation}
\label{couplazzo3}
\frac{ \partial \mathcal E }{ \partial \bm{\varepsilon}_{\textup{p}} }[\bm{\varepsilon} - \bm{\varepsilon}_{\textup{p}}]
= \tr( \bm{\varepsilon} - \bm{\varepsilon}_{\textup{p}} ) \, \frac{ \partial \lambda }{ \partial \bm{\varepsilon}_{\textup{p}} }
\otimes \bm{I}
+ 2 \frac{\partial \mu}{ \partial \bm{\varepsilon}_{\textup{p}} }
\otimes \left( \bm{\varepsilon} - \bm{\varepsilon}_{\textup{p}} \right) \, .
\end{equation}

A process of isostatic pressing is now considered of a ceramic powder subject to increasing pressure $p$.
In terms of finite (not incremental) quantities, the volumetric strain $e_{\textup{v}}$ is related to the stress variable $p$ through
\begin{equation}
\label{vacca}
e_{\textup{v}} = -\frac{p}{ K(e_{\textup{v}}^{\textup{p}}) } + e_{\textup{v}}^{\textup{p}} \, ,
\end{equation}
where $K(e_{\textup{v}}^{\textup{p}})$ is the elastic bulk modulus, dependent on the plastic volumetric strain.
This dependence can be immediately obtained from equation~\eqref{wlf} when the following relation between the density and the plastic volumetric strain is taken into account
\begin{equation}
\rho = \frac{\rho_{0}}{ 1+e_{\textup{v}}^{\textup{p}} },
\end{equation}
where $\rho_{0}$ is the initial density.

If the Cooper and Eaton~\cite{cooper} isostatic compaction model is assumed (equation~\eqref{coop}), the plastic volumetric strain $e_{\textup{v}}^{\textup{p}}$ is related to the hydrostatic compaction pressure $p$ as
\begin{equation}
\label{boia}
e_{\textup{v}}^{\textup{p}} = \left( \frac{ \rho_{0} }{ \rho_{\infty} } - 1 \right) \exp \left( -\frac{b}{p} \right) \, ,
\end{equation}
while, if the simpler exponential law~\eqref{exp} is assumed, the following expression is obtained
\begin{equation}
\label{stokka}
e_{\textup{v}}^{\textup{p}} = \frac{\rho_{0}}{ \rho_{\infty} - (\rho_{\infty} - \rho_{0})\exp(-bp) } - 1 \, .
\end{equation}

\begin{figure}[!htb]
\centering
\includegraphics[width=17cm,keepaspectratio]{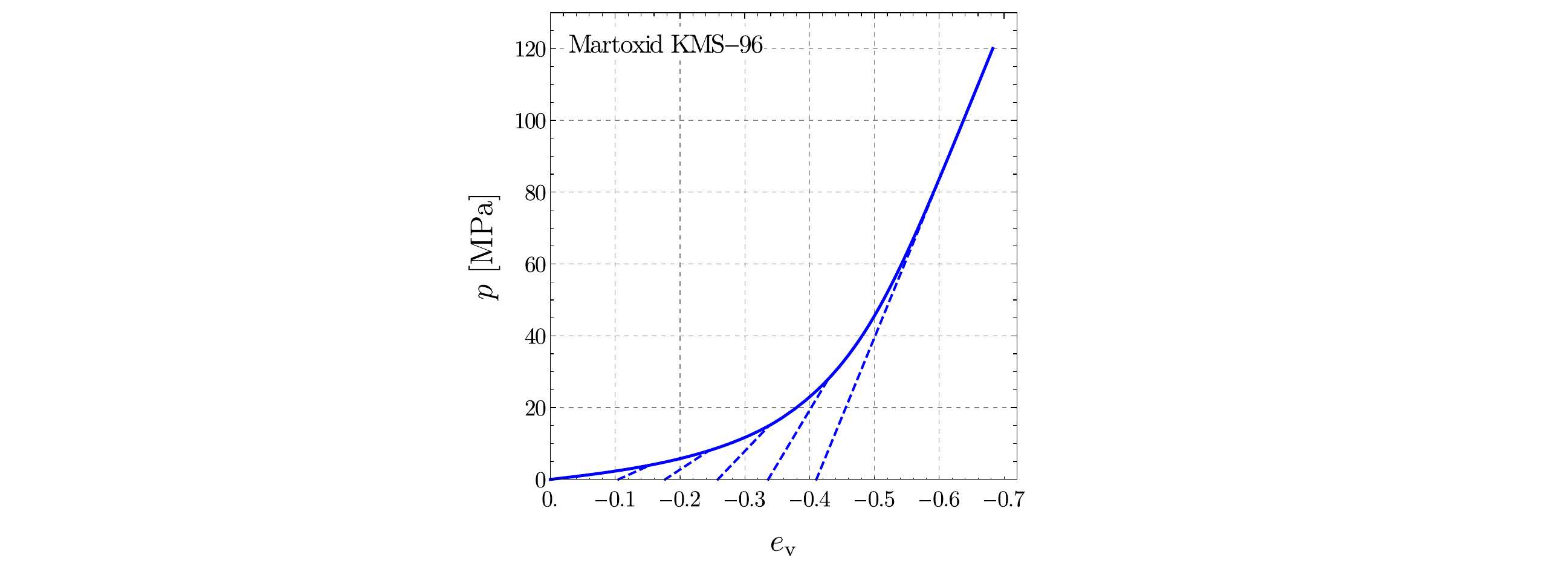}
\caption{Predicted isostatic forming response at increasing pressure $p$, with unloading for an alumina powder (Martoxid KMS-96).
Note the progressive increase in stiffness of the elastic response (visible at unloading), due to elastoplastic coupling.}
\label{stiffening}
\end{figure}

Using the law~(\ref{stokka}) with the values of the constants $\rho_{0}$, $\rho_{\infty}$, and $b$ listed in figure~\ref{densityfott} (upper part, on the right) into equation~\eqref{vacca},
the isostatic forming response of the material can be predicted. This prediction is plotted in figure~\ref{stiffening}, which provides a good qualitative simulation of a compaction curve for a ceramic powder, thus validating the elastoplastic coupling constitutive modelling. 

\section[Conclusions]{Conclusions}

The variation of the elastic properties (elastic modulus and Poisson's ratio) of green bodies after compaction at different pressures has been investigated using ultrasonic P and S waves propagation, through the introduction of a specifically designed testing protocol.
The variation has been found significant and definitely non-negligible in the modelling and simulation of the behaviour of ceramic powders during forming.
Three different laws have been proposed, all of which provide a good interpolation of the data in terms of dependence of the P and S wave speeds and of the elastic constants on the forming pressure.
Plotted as functions of the density, the variations of all these quantities become linear, even if it is clear that this linearity only holds true at medium and high density, but is definitely false at low densities (not investigated in the present article). The laws providing the elastic constants as functions of the density have been shown to be fully compatible with the concept of elastoplastic coupling and an example has been provided of simulation of a compaction curve. 
The validation of the experimental results through a micromechanical model based on the deformation of a packaging of elastoplastic cylinders and spheres is deferred to Part~II of this study.

It should be highlighted in conclusion that the proposed experimental protocol could easily be modified to produce a robust, validatable quality control test protocol in an industrial production environment.
This protocol could be inserted at a point in the production of ceramics before the firing process.
This would reduce energy consumption caused by the firing of non conforming products.

\section*{Acknowledgments}
The authors gratefully acknowledge financial support from  the European FP7  programs INTERCER-2 PIAP-GA-2011-286110-INTERCER2 (L.P.A., D.C., D.M., and Z.V.),
ERC-2013-ADG-340561-INSTABILITIES (D.B.), and FP7-PEOPLE-2013-ITN-PITN-GA-2013-606878-CERMAT2 (A.P.).

\end{document}